\documentclass[hidelinks,10pt]{article}
\usepackage{arxiv}
\usepackage{nicefrac}      
\usepackage{microtype}      % microtypography
\usepackage{booktabs}
\usepackage{doi}
\usepackage{adjustbox}
\usepackage{amsmath}
\usepackage{amssymb}
\usepackage{float}
\usepackage[T1]{fontenc}
\usepackage{graphicx}
\usepackage{hhline}
\usepackage[utf8]{inputenc}
\usepackage{mathtools}
\usepackage{multicol}
\usepackage{multirow}
\usepackage{wasysym}
\usepackage{hyperref}
\usepackage[round]{natbib}
\usepackage{geometry}
\usepackage{longtable}
\usepackage{enumitem} % For continuous numbering

\usepackage{setspace}
\title{Neighborhood Disparities in Smart City Service Adoption}

\author{
    Shahaf Donio \\
    IWiT Lab \\ Department of Industrial Engineering \\ Tel Aviv University
    \and 
    Eran Toch\thanks{Email: erant@tauex.tau.ac.il} \\
    IWiT Lab \\ Department of Industrial Engineering \\ Tel Aviv University 
}

\renewcommand{\headeright}{}
\renewcommand{\shorttitle}{Neighborhood Disparities in Smart City Service Adoption}

\hypersetup{
pdftitle={Neighborhood Disparities in Smart City Adoption},
pdfsubject={}, % Change Subject here
pdfauthor={},
pdfkeywords={},
}

\begin{document}
\maketitle

\begin{abstract}
While local governments have invested heavily in smart city infrastructure, significant disparities in adopting these services remain in urban areas. The success of many user-facing smart city technologies requires understanding barriers to adoption, including persistent inequalities in urban areas. An analysis of a random sample telephone survey (n=489) in four neighborhoods of Tel Aviv merged with digital municipal services usage data found that neighborhood residency influences the reasons why residents adopt resident-facing smart city services, as well as individual-level factors. Structured Equation Modeling shows that neighborhood residency is related to digital proficiency and privacy perceptions beyond demographic factors and that those influence the adoption of smart-city services. We summarize the paper by discussing why and how place effects must be considered in further research in smart cities and the study and mitigation of digital inequality.
\end{abstract}

\section{Introduction}

In recent years, many cities worldwide have deployed new technologies and embedded information systems in urban environments. Technologies such as traffic sensors, wireless communication networks, intelligent cameras, and mobile municipal apps are being increasingly adopted in urban environments today. Existing definitions of "smart cities" point to using technologies in urban environments \citep{singh2022decade,morvaj2011demonstrating,batty2012smart,camero2019smart}. Smart cities also represent a social and organizational process in which municipalities and urban environments change by collecting and analyzing data \citep{hollands2015critical,hatuka2018political,gaffney2018smarter}. Through the last decade, cities have been embedding computing and communication technologies at an increasing pace \citet{de2024smart}. This considerable investment, funded chiefly by taxpayers’ money, raises important questions about how innovative city technologies address and fulfill citizens’ needs.

The debate regarding the ability of smart cities to fulfill the pledges made by local authorities and tech firms is vibrant and multi-dimensional. Smart cities are celebrated as a new model that has the potential to address urban challenges. This is accomplished through enhancing cost-effectiveness in city management \cite{shamsuddin2021just}, laying the groundwork for infrastructures that pave the way for new urban activities \cite{krishnamurthi2019innovation}, and aiding in the cultivation of an environment conducive to innovation \cite{ooms2020ecosystems}. More broadly, smart cities may contribute to the local economy and thus may lead to lower levels of urban income inequality \citep{caragliu2022smart}.

More critical perspectives of smart cities point to concerns for privacy invasion \citep{martinez2013pursuit,kitchin2014real,rizi2022systematic,curzon2019survey}, susceptibility to cyber vulnerabilities  \cite{zhang2017security,laufs2020security}, or even to the potential damage that can caused by buggy software \cite{kitchin2014real}. Particular concerns relate to the potential of urban digitization in broadening city inequality. Scholars have raised concerns regarding the exclusionary nature of smart cities, particularly regarding accessibility to services for marginalized communities, such as older people \citep{calzada2015unplugging}. Smart cities may enable municipal governments and tech companies to amass and centralize power \cite{gaffney2018smarter,sadowski2021owns,cardullo2019being,de2024smart}. More profoundly, smart cities may shift the basic relationships between cities and their citizens to resemble more the relationship between private companies and their customers \cite{filion2023urban,das2024exploring,johnson2020type}. 

While implementing innovative city initiatives varies across different countries, cities, and technological objectives \citep{singh2022decade}, there is a unifying core principle underlying all of them - the emphasis on resident-centric government services. In contrast to certain government services like tax payment systems, which may be mandatory, many smart city services directed at residents are optional. The success of these services hinges on the residents' inclination to understand, use, and promote them \cite{dirsehan2022smart,chatterjee2018effects}. The extent of acceptance, dissemination, and deployment of e-government initiatives are intrinsically tied to the residents' propensity to avail themselves of these services.

However, the e-government adoption models presented in existing literature \cite{shareef2011government} do not consider some of the most critical urban structures of the city. While empirical studies grounded in these models have examined demographic factors \cite{yeh2017effects,zheng2014urban,chatterjee2018success,inkinen2018variations,habib2023factors,deng2023exploring,marimuthu2022integrating}, they have not taken into account the direct effect of neighborhoods. The closest work to our own is by \citet{inkinen2018variations}, which analyzed three neighborhoods, but the work did not evaluate the contribution of neighborhoods in comparison to the demographic properties of the residents. characteristics of the residents. Nevertheless, it is acknowledged that neighborhoods and locales hold significant relevance in urban informatics \cite{cranshaw2012livehoods,zhang2019public,hatuka2024conceptual} and that neighborhood-level factors influence the adoption of technologies such as home Internet access \cite{mossberger2007digital}. Therefore, this paper asks whether and how the adoption of innovative city initiatives is influenced by the neighborhoods in which residents reside. 

In this study, we explore the concept of neighborhoods as entities that might embody demographic traits associated with the embrace of smart city technologies. We build upon the e-Government Adoption Model (GAM) by tailoring it to the technology adoption process among city inhabitants. Employing a survey conducted across four distinct neighborhoods in Tel Aviv, we discern varying adoption patterns consistent with the diverse demographic profiles of the neighborhoods. Our analysis leads us to conclude that all three of the aforementioned intermediary variables play a pivotal role in the outcomes of technology adoption. We culminate the study by proposing guidelines for designing and implementing technologies that aim to bolster municipal services and cater to the needs of city residents. This aims to ensure that smart city initiatives are effective and aligned with the varying demographic dynamics.

\section{Theoretical Background}

The success and operation of smart cities are inherently related to the adoption of smart city technologies by residents and visitors. The user adoption model of government digital services was documented in several models. The e-Government Adoption Model (GAM) \citet{shareef2011government} combines several technological diffusion theories and was later extended to mobile government services \citet{shareef2016service}. Several aspects are essential in determining the adoption of technologies: Awareness contributes to the willingness to adopt new technology, and end-user transformation is a gradual and time-consuming process. 

Some smart city technologies can be very intrusive towards residents' privacy, potentially compromising their interaction with digital services, the traces that residents leave through various sensing technologies \citep{martinez2013pursuit}. As smart cities are built as an infrastructure, they can lead to over-collection driven by the many sensors, services, and interfaces \citep{li2015privacy}. The centralized architecture of many smart city infrastructure products can lead to concentrating personal information, which makes it more vulnerable to privacy and security risks \citep{elmaghraby2014cyber}. While privacy and security were found to be predictive of adoption in many smart city adoption models \cite{dirsehan2022smart} and in e-government adoption models \citep{shareef2011government}, they were not evaluated about places across the city.

When analyzing heterogeneous urban experiences in different neighborhoods, the associated factors cannot be mapped just to demographic characteristics such as ethnicity or income \citep{sharkey2014and}. This is also true for digital experiences, shaped by people's neighborhoods \citep{hatuka2024conceptual}. Neighborhoods are created through diverse and dynamic processes, in which particular groups are attracted to specific neighborhoods \citep{sampson2008neighborhood} and develop specific norms and culture \citep{bader2016fragmented}. The process in which these disparities can be explained by the degree of homogeneity in the neighborhoods, which depends on the social dynamics in the neighborhood \citep{tulin2021same}. Once individuals have settled into a neighborhood, the neighborhood composition can directly induce network composition.

Contemporary approaches to the study of the digital divide go beyond basic access to technology to psychological and social reasons that prevent marginalized communities from correctly accessing smart city services \citep{partridge2004developing,effing2016social}. Before individuals can adopt internet technology that allows them to use the internet effectively, the technology must be available in the form of infrastructure that is accessible and usable in their cultural context \citep{fakude2023role}. Barriers to accessing digital technologies include norms of technology use \citep{ragnedda2013digital} and proficiency in Internet skills  \cite{hargittai2008digital,murray2014unraveling,van2016development}. The adoption of ICT is mediated by various social factors, such as age \citep{neves2013coming}, gender \citep{cooper2006digital}, income \citep{patria2020impact}, ethnicity \cite{walker2020exploring}, religious beliefs \citep{tsuria2020digital}, and language \citep{wolk2004effects}. However, analyzing the adoption of technologies such as Broadband access points to social and economic aspects of the neighborhood of determining meaningful access to technology \citet{ragnedda2013digital}. Neighborhood-level factors may present additional constraints on access to technology beyond individual-level characteristics such as education, age, ethnicity, occupation, and ethnicity \citep{,mossberger2012unraveling,kaplan2012prospects}. The effect might be even more pronounced in innovative city services, as they are built and designed according to specific forms of urban lifestyles.

\section{Research Scope and Hypotheses}

\begin{figure}[!htbp]
\includegraphics[width=\linewidth]{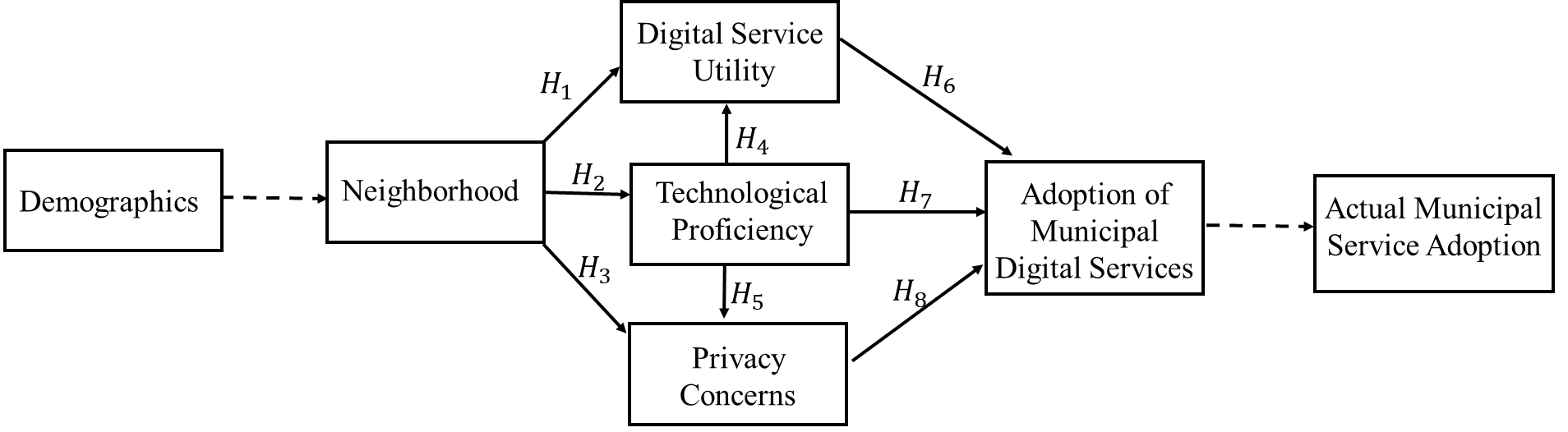}
\caption{Research model and hypotheses}
\label{fig:research_framework}
\end{figure}

In this section, we present a model of user-facing smart city technologies. The model is based on the hypothesis that the neighborhood represents a group of people with the same beliefs and abilities and that, therefore, their technology skills are influenced. Additional concepts and capabilities (smart city utility and perceived trust) will ultimately affect the usage of smart city services. Conceptual definitions, as well as theoretical arguments for the proposed relationship, are developed below. 

Our basic observation is that the neighborhood represents people with similar characteristics. Therefore, the neighborhood, as a group of people who have some level of homogeneity, which is characterized by similar and common properties, may be a group with a similar affinity to technology and a similar worldview and imagination in terms of its members’ different approaches to information privacy and information security. This leads us to posit the following hypotheses:

\section{Hypotheses Development}

The adoption of smart city technologies is influenced by various factors, including social, cultural, and technological aspects that vary across neighborhoods. Previous research has highlighted the importance of understanding how urban environments shape technology adoption, emphasizing the interplay between individual characteristics and neighborhood-level dynamics \citep{sampson2008neighborhood, ragnedda2013digital}. Based on these insights, we propose the following hypotheses:

Neighborhoods, as social and cultural units, shape collective norms, beliefs, and resource access, influencing residents’ perceptions of technology. The specific utilities that residents derive from smart city technologies are, therefore, likely to vary across neighborhoods due to differences in socioeconomic status, education, and infrastructure. Thus, we hypothesize that:

\section{Hypotheses Development}

A combination of individual, social, and contextual factors shapes the adoption of smart city technologies. Understanding these dynamics requires integrating insights from urban sociology, technology adoption theories, and privacy studies. Prior literature highlights the importance of spatial and social structures, particularly neighborhoods, in influencing the adoption and utility of digital technologies \citep{sampson2008neighborhood, ragnedda2013digital}. The following hypotheses are developed to explore these relationships in detail.

Our first set of hypotheses relies on the observation that neighborhoods are not just physical spaces but are also defined by their social, cultural, and economic compositions. These factors influence how residents perceive and engage with technology. For instance, neighborhoods with higher socioeconomic status may provide better access to technological infrastructure and support systems, enabling residents to derive higher utility from smart city technologies \citep{caragliu2022smart}. In contrast, neighborhoods with lower socioeconomic status may lack such resources, reducing the perceived utility of general digital services (such as social networks and bill payment). In these hypotheses, we assume that neighborhood residency will explain more variability than demographics alone. Thus, we hypothesize:

\textbf{H\textsubscript{1}}: The neighborhood residency is associated with the individual's perceived utility of the digital services.

Access to education, training, and resources varies significantly across neighborhoods, creating disparities in technological skills. Affluent neighborhoods are more likely to have higher levels of digital literacy due to better access to schools \citep{ragnedda2013digital}. On the other hand, residents of underprivileged neighborhoods may face barriers such as limited infrastructure and fewer opportunities for skill development. Based on this, we hypothesize:

\textbf{H\textsubscript{2}}: The neighborhood residency is associated with the individual's technological proficiency.

Privacy concerns, while necessary for trust in digital systems, are shaped by residents’ awareness and experiences. Marginalized communities often lack the resources to understand and address privacy risks, leading to uneven privacy skills \citep{gangadharan2017downside}. This disparity is further exacerbated by systemic inequalities, where some neighborhoods are more likely to face surveillance or data misuse. As privacy concerns are inherently tied to technological awareness and contextual experiences, we hypothesize:

\textbf{H\textsubscript{3}}: Neighborhood residency is associated with the privacy concerns of the residents.

An individual's technological proficiency influences the perceived utility of digital services. Users more skilled in navigating technology are better equipped to derive value from digital services, as they can utilize advanced features and solve common issues independently \citep{hargittai2004internet, sengpiel2015validation}. This ability to maximize utility increases satisfaction and encourages continued use. Therefore, we hypothesize:

\textbf{H\textsubscript{4}}: Technological proficiency is positively associated with digital service utility.

Technological proficiency also shapes how individuals perceive and manage privacy risks. Digitally skilled users are more likely to be aware of privacy implications, such as data breaches and surveillance, and take proactive steps to protect their information \citep{park2013digital, hargittai2008digital}. This heightened awareness, while beneficial, can also lead to more significant concerns about privacy. Thus, we hypothesize:

\textbf{H\textsubscript{5}}: Technological proficiency is positively associated with privacy concerns.

Attitudes towards digital service utility are a strong predictor of technology adoption. When residents perceive digital services as applicable and aligned with their needs, they are more likely to integrate them into their daily routines \citep{shareef2011government}. Therefore, we hypothesize:

\textbf{H\textsubscript{6}}: Attitudes towards digital service utility are positively associated with a higher willingness to adopt user-facing smart city services.

Technological proficiency enables users to engage with smart city initiatives more effectively, making it a critical factor for adoption. Proficient users can navigate digital interfaces, troubleshoot problems, and understand the benefits of smart city services, leading to higher adoption rates \citep{stern2009digital}. Consequently, we hypothesize:

\textbf{H\textsubscript{7}}: Technological proficiency is positively associated with a higher willingness to adopt user-facing smart city services.

Privacy concerns can act as both a motivator and a barrier to adoption. While addressing privacy risks is crucial for building trust, heightened concerns may deter individuals from using digital services \citep{dinev2006extended, gangadharan2017downside}. Residents who have higher concerns and lower trust in ICT may avoid adopting these services unless adequate safeguards are in place. Based on this, we hypothesize:

\textbf{H\textsubscript{8}}: Privacy concerns are negatively associated with a higher willingness to adopt user-facing smart city services.

\section{Methodology}

\subsection{Procedure and participants}

\begin{figure}[ht]
\centering
\includegraphics[width=12cm]{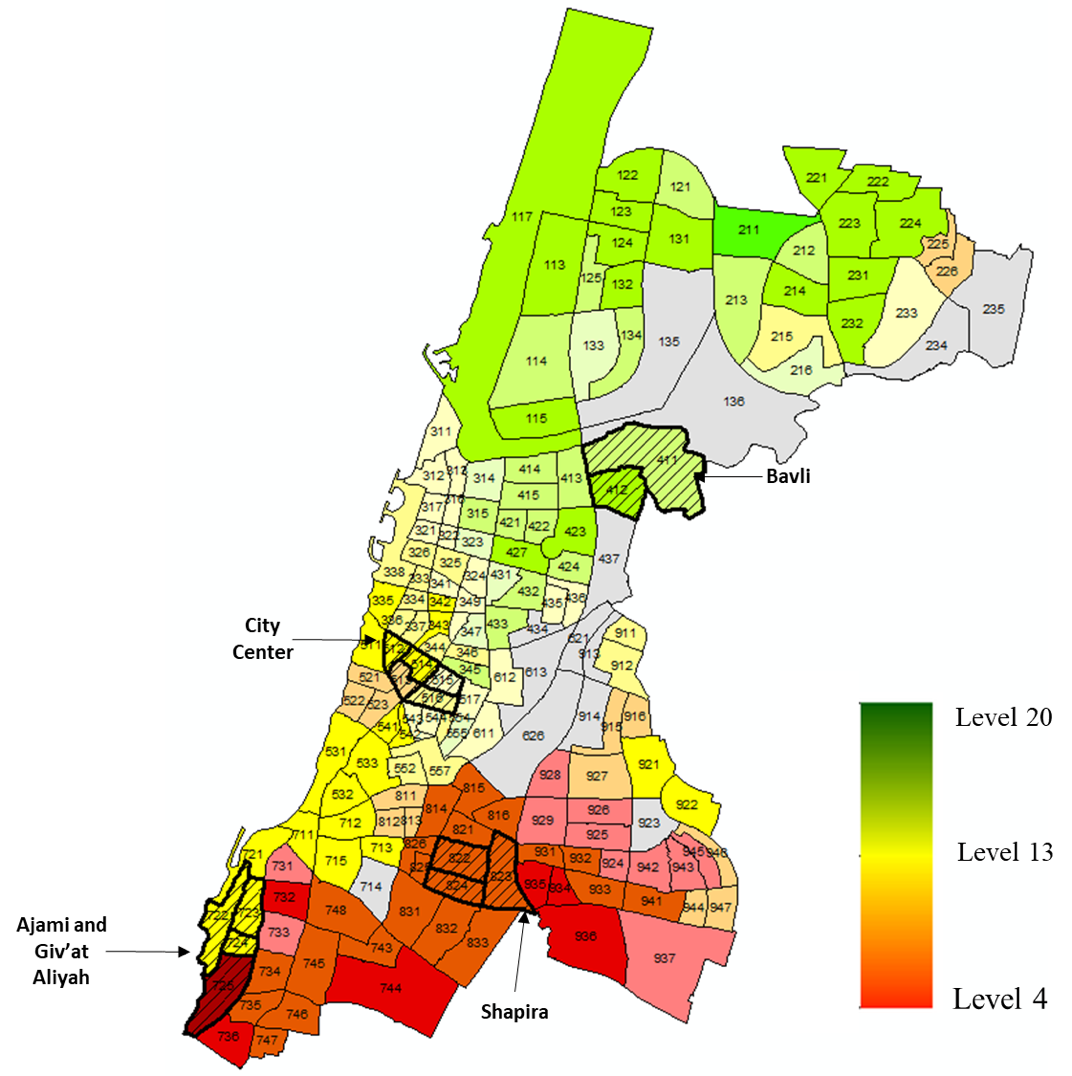}
\caption{Tel Aviv-Yafo city map by statistical area divisions, with the sampled neighborhoods highlighted. Color represents the socioeconomic index (according to Israel's Central Bureau of Statistics): red represents low socioeconomic status, and green represents high socioeconomic status.   
}
\label{fig:telaviv}
\end{figure}

The survey was conducted on a representative telephone sample using cluster sampling techniques in four neighborhoods selected in Tel Aviv, Israel: Bavli, City Center, Ajami and Giv'at Aliyah, and Shapira. The neighborhoods, visualized in Figure \ref{fig:telaviv}, were chosen by their geographical location based on what we knew from previous information and our media research, indicating that each neighborhood has different characteristics. Each selected neighborhood represents a different group of people with different background characteristics. The neighborhoods in the south represent a population with a weaker economic background, with the leading language in each neighborhood being Hebrew or Arabic. In addition, we chose two northern neighborhoods with stronger economic backgrounds, with the distinguishing feature that one of the neighborhoods is considered a young people's neighborhood and the other an older people's neighborhood.

The survey was executed by B. I. and Lucille Cohen Institute for Public Opinion Research\footnote{\url{https://en-social-sciences.tau.ac.il/bicohen}}, an institute that conducts surveys for researchers for academic institutions. In each neighborhood, we defined the number of people who would answer the questionnaire, and the surveyors conducted the survey until we received enough samples. In each neighborhood, 123 residents responded to the survey. 

The survey was conducted over a three-month period (November 2016–January 2017). To distribute the questionnaire, we sampled 11 different age groups (between 18 to 70+), with at least 15 residents from each group. Furthermore, we sampled 50$\%$ of the residents to be male and 50$\%$ of the residents to be female and 8 different education groups, with at least 15 residents from each group. To ask the participants questions in their mother tongue, the questionnaire was asked in Hebrew or Arabic. The full breakdown of demographic data per neighborhood is presented in Table \ref{tab:demographics}. The response rate in each neighborhood was: Bavli (22$\%$), City Center (21$\%$), Shapira (29$\%$), and Ajami (29$\%$).The population that was not surveyed includes homeless people, refugees, people who do not have phone/cellular phones and residents under 18. We received a total of 490 questionnaires, and one returned empty. Therefore, we had 489 eligible response questionnaires. 

\begin{longtable}{|l|l|llll|}
\caption{Demographic Data by Neighborhood} \label{tab:demographics} \\ \hline
\textbf{Variable} & \textbf{Categories} & \multicolumn{4}{l|}{\textbf{Neighborhood}} \\ \hline
\endfirsthead
\hline
\textbf{Variable} & \textbf{Categories} & \multicolumn{4}{l|}{\textbf{Neighborhood}} \\ \hline
\endhead
\hline
\endfoot 
\multirow{4}{*}{Age}         &  & \multicolumn{1}{l|}{Bavli} & \multicolumn{1}{l|}{City Center} & \multicolumn{1}{l|}{Shapira} & Ajami \\ \cline{2-6} 
                          & Mean                             & \multicolumn{1}{l|}{50.23} & \multicolumn{1}{l|}{44.21}       & \multicolumn{1}{l|}{46.17}   & 47.17 \\ \cline{2-6} 
                          & Standard Deviation               & \multicolumn{1}{l|}{18.84} & \multicolumn{1}{l|}{17.09}       & \multicolumn{1}{l|}{16.44}   & 17.81 \\ \cline{2-6} 
                          & Other                            & \multicolumn{1}{l|}{14}    & \multicolumn{1}{l|}{3}           & \multicolumn{1}{l|}{18}      & 10    \\ \hline
\multirow{4}{*}{Gender}         &  & \multicolumn{1}{l|}{Bavli} & \multicolumn{1}{l|}{City Center} & \multicolumn{1}{l|}{Shapira} & Ajami \\ \cline{2-6} 
                          & Male                             & \multicolumn{1}{l|}{51}    & \multicolumn{1}{l|}{59}          & \multicolumn{1}{l|}{51}      & 51    \\ \cline{2-6} 
                          & Female                           & \multicolumn{1}{l|}{72}    & \multicolumn{1}{l|}{63}          & \multicolumn{1}{l|}{70}      & 73    \\ \cline{2-6} 
                          & Other                            & \multicolumn{1}{l|}{0}     & \multicolumn{1}{l|}{0}           & \multicolumn{1}{l|}{0}       & 0     \\ \hline
\multirow{10}{*}{Education}     &  & \multicolumn{1}{l|}{Bavli} & \multicolumn{1}{l|}{Center}      & \multicolumn{1}{l|}{Shapira} & Ajami \\ \cline{2-6} 
                          & Elementary school (A1)           & \multicolumn{1}{l|}{0}     & \multicolumn{1}{l|}{0}           & \multicolumn{1}{l|}{7}       & 15    \\ \cline{2-6} 
                          & partial high school (A2)         & \multicolumn{1}{l|}{0}     & \multicolumn{1}{l|}{1}           & \multicolumn{1}{l|}{11}      & 20    \\ \cline{2-6} 
                          & high school without diploma (A3) & \multicolumn{1}{l|}{5}     & \multicolumn{1}{l|}{4}           & \multicolumn{1}{l|}{27}      & 21    \\ \cline{2-6} 
                          & high school with diploma (A4)    & \multicolumn{1}{l|}{18}    & \multicolumn{1}{l|}{21}          & \multicolumn{1}{l|}{34}      & 23    \\ \cline{2-6} 
                          & formal education (A5)            & \multicolumn{1}{l|}{12}    & \multicolumn{1}{l|}{18}          & \multicolumn{1}{l|}{14}      & 10    \\ \cline{2-6} 
                          & partial academic degree (A6)     & \multicolumn{1}{l|}{2}     & \multicolumn{1}{l|}{4}           & \multicolumn{1}{l|}{6}       & 6     \\ \cline{2-6} 
                          & bachelor's degree (A7)           & \multicolumn{1}{l|}{45}    & \multicolumn{1}{l|}{49}          & \multicolumn{1}{l|}{17}      & 19    \\ \cline{2-6} 
                          & Master degree and above (A8)     & \multicolumn{1}{l|}{41}    & \multicolumn{1}{l|}{25}          & \multicolumn{1}{l|}{5}       & 9     \\ \cline{2-6} 
                          & Other                            & \multicolumn{1}{l|}{0}     & \multicolumn{1}{l|}{0}           & \multicolumn{1}{l|}{0}       & 1     \\ \hline
\multirow{8}{*}{Monthly income} &  & \multicolumn{1}{l|}{Bavli} & \multicolumn{1}{l|}{Center}      & \multicolumn{1}{l|}{Shapira} & Ajami \\ \cline{2-6} 
                          & Less than 1,200\$ (B1)           & \multicolumn{1}{l|}{3}     & \multicolumn{1}{l|}{7}           & \multicolumn{1}{l|}{28}      & 26    \\ \cline{2-6} 
                          & 1,201$ - 2,300$ (B2)             & \multicolumn{1}{l|}{13}    & \multicolumn{1}{l|}{14}          & \multicolumn{1}{l|}{31}      & 31    \\ \cline{2-6} 
                          & 2,301$ - 3,400$ (B3)             & \multicolumn{1}{l|}{17}    & \multicolumn{1}{l|}{16}          & \multicolumn{1}{l|}{26}      & 10    \\ \cline{2-6} 
                          & 3,401$ - 4,700$ (B4)             & \multicolumn{1}{l|}{13}    & \multicolumn{1}{l|}{16}          & \multicolumn{1}{l|}{4}       & 3     \\ \cline{2-6} 
                          & 4,701$ - 6,000$ (B5)             & \multicolumn{1}{l|}{10}    & \multicolumn{1}{l|}{10}          & \multicolumn{1}{l|}{3}       & 2     \\ \cline{2-6} 
                          & 6,001\$ or more (B6)             & \multicolumn{1}{l|}{18}    & \multicolumn{1}{l|}{15}          & \multicolumn{1}{l|}{0}       & 2     \\ \cline{2-6} 
                          & Other                            & \multicolumn{1}{l|}{49}    & \multicolumn{1}{l|}{44}          & \multicolumn{1}{l|}{29}      & 50    \\ \hline
\multirow{5}{*}{Religion} &                                  & \multicolumn{1}{l|}{Bavli} & \multicolumn{1}{l|}{Center}      & \multicolumn{1}{l|}{Shapira} & Ajami \\ \cline{2-6} 
                          & Jewish                           & \multicolumn{1}{l|}{121}   & \multicolumn{1}{l|}{117}         & \multicolumn{1}{l|}{115}     & 18    \\ \cline{2-6} 
                          & Muslim                           & \multicolumn{1}{l|}{0}     & \multicolumn{1}{l|}{1}           & \multicolumn{1}{l|}{0}       & 75    \\ \cline{2-6} 
                          & Christian                        & \multicolumn{1}{l|}{0}     & \multicolumn{1}{l|}{1}           & \multicolumn{1}{l|}{2}       & 28    \\ \cline{2-6} 
                          & Other                            & \multicolumn{1}{l|}{2}     & \multicolumn{1}{l|}{3}           & \multicolumn{1}{l|}{4}       & 3     \\ \hline
\multirow{7}{*}{Level of Religiousness}    &                                  & \multicolumn{1}{l|}{Bavli} & \multicolumn{1}{l|}{Center}      & \multicolumn{1}{l|}{Shapira} & Ajami \\ \cline{2-6} 
                          & Secular (C1)                     & \multicolumn{1}{l|}{93}    & \multicolumn{1}{l|}{100}         & \multicolumn{1}{l|}{36}      & 29    \\ \cline{2-6} 
                          & Traditional (C2)                 & \multicolumn{1}{l|}{15}    & \multicolumn{1}{l|}{9}           & \multicolumn{1}{l|}{22}      & 16    \\ \cline{2-6} 
                          & Traditional-Orthodox (C3)        & \multicolumn{1}{l|}{9}     & \multicolumn{1}{l|}{2}           & \multicolumn{1}{l|}{28}      & 37    \\ \cline{2-6} 
                          & Orthodox (C4)                    & \multicolumn{1}{l|}{2}     & \multicolumn{1}{l|}{3}           & \multicolumn{1}{l|}{20}      & 22    \\ \cline{2-6} 
                          & Ultra – Orthodox (C5)            & \multicolumn{1}{l|}{1}     & \multicolumn{1}{l|}{2}           & \multicolumn{1}{l|}{7}       & 11    \\ \cline{2-6} 
                          & Other                            & \multicolumn{1}{l|}{3}     & \multicolumn{1}{l|}{6}           & \multicolumn{1}{l|}{8}       & 9     \\ \hline
\end{longtable}

\subsection{Sample Validity}
To examine the validity of our sample we test whether it is chosen in each region indeed represented the population of the region, we compared the research data with the data obtained from the Tel Aviv central bureau of statistics. For each neighborhood and each region, we examined the proportion of the population by two variables: age and gender. To test the homogeneity of proportions, we used the $\lambda$\textsuperscript{2}–test for the equality of proportion. The hypothesis we checked states that the distribution of the age groups for each neighborhood is the same as the neighborhood region. After we collected the data, we used the percentages of each population group to compare the distributions. In the same way, we tested for gender. 

Table \ref{tab:demographics_validation} shows the demographic results of the entire questionnaire by division for each neighborhood and per demographic question. We also present the calculation of the number and percentages for each population group and the $\lambda$\textsuperscript{2}–test p-value result for each of the 8 tests we conducted. For all of the tests that we performed, we accepted the hypothesis, meaning that we do not have strong evidence that the population that we tested was distributed differently from the population of the region, which means that the neighborhood sample that we tested indeed represented the region. 

\subsection{Tel Aviv Smart City Services}

Tel Aviv is a relevant case study for analyzing smart city adoption for several reasons. First, it has relatively advanced resident-facing smart city technologies \citep{weinstein2017digi,feder2016well,madakam2018internet}. Tel Aviv provides access to broadband connectivity and high-speed mobile data connections throughout the city. Tel Aviv is the second-largest city in Israel and the most densely populated area in the country. The diversity and heterogeneity of the Tel Aviv population is another strong aspect of the case study. The city is inhabited by young people (21.7$\%$), adults (58.6$\%$), elderly people (19.7$\%$), and Jewish (91.3$\%$) and Arab (8.3$\%$) residents; by secular and religious residents; by the rich and the poor; and by women (50.8$\%$) and men (49.2$\%$)\footnote{Statistics Provided by the Center for Economic and Social Research of Tel-Aviv-Yafo Municipality}.

The main digital service we have measured is Tel Aviv's flagship project, the DigiTel project \citep{weinstein2017digi}. The DigiTel system, established in 2013, offers several services for residents, such as discounts for theaters, museums, and activities throughout the city; a bicycle subscription and car-sharing system; and online services for paying bills and issues related to compulsory payments \citep{steinmetz2019tel}. The DigiTel service is available through the internet, mobile, and phone interfaces, and some physical services are accessible through a physical card. The services allow residents to engage with the municipality—residents can locate events, report hazards around the city, follow up on their handling, and follow events \citep{weinstein2020humanize}. The city also actively employs social media as an essential platform for involving the public in municipal news and community improvements. Today, Tel Aviv has expanded the DigiTel platform, including Digi-Dog for dog owners and Digi-Taf (young children in Hebrew), serving different users across the city.

\subsection{Measures}

\paragraph{Dependent Variable}
\textbf{Adoption of Municipal Digital Services (AoMDS):} This construct reflects residents' engagement with municipal digital services, capturing the frequency of their interactions based on \citet{shareef2011government}. It includes usage patterns of platforms like DigiTel for tasks such as payments, registration, and accessing municipal benefits. Items within this measure assess not only the extent but also the purpose of service use, providing a comprehensive view of adoption behaviors. 

To provide additional external validity to the adoption scale, we compared the aggregated neighborhood-level variable to two other variables:
\begin{enumerate}[label=(\alph*)]
    \item The number of new residents registered for the DigiTel service according to Tel Aviv Municipality.
    \item The number of residents’ geo-located interactions with Tel Aviv's Facebook page. 
\end{enumerate}

\paragraph{Independent Variables}
\begin{itemize}
    \item \textbf{Neighborhood Residency:} The categorical variable representing the resident’s neighborhood (Bavli, City Center, Shapira, Ajami). It serves as a primary contextual factor influencing technological adoption, reflecting spatial and socioeconomic disparities.

    \item \textbf{Technological Proficiency (TP):} This variable measures the resident's skill level with technology, including familiarity and frequency of use of devices like smartphones and computers, as well as the ability to download and use applications. Items assess both basic and advanced digital skills, based on \citet{murray2014unraveling,van2016development}.

    \item \textbf{Digital Service Utility (DSU):} This construct evaluates the perceived benefits and functionality of digital services from the user’s perspective. It includes components such as communication, civic engagement, and interactions with public authorities.

    \item \textbf{Privacy Concerns (PC):} This variable captures individual apprehensions about the collection and use of personal information by municipal digital systems. It assesses concerns related to data collection on residents' activities and the security of municipal services. Questions are based on \citet{chang2018role} and on \citet{kim2010empirical}.

    \item \textbf{Demographics:} Several demographic factors (age, gender, education, income, and language) serve as controls, allowing the study to isolate the impact of neighborhood-level influences on the dependent variable.
\end{itemize}
The full list of questions can be seen in Table \ref{tab:items}.

\subsection{Factor Analysis}

We first conduct a Confirmatory Factor Analysis (CFA) on the preliminary 16 scale items, measuring the factor loading about the latent variables. The full results are available in Table \ref{tab:factor_loadings}. We calculated the factor loading on the 16 measuring items, removing items lower than the cutoff value of 0.05 \citep{rummel1988applied}. After removing one item, we found that 3 constructs with 13  items could be retained. To support the factor definition, we also examined the correlation matrix analyzed by the CFA. Next, we verified the correlation between all items.  All the observed Pearson’s $r$ values corresponded to less than 5$\%$ common variance. 

To measure the fit scores of the retained measurements, we used the coefficient Cohen alpha test. The reliability scores for all the final exogenous and endogenous variables range from 0.65 to 0.84. This score suggests an acceptable internal consistency among the items in each dimension \citep{fabrigar1999evaluating}. We found that 3 variables, perceived trust (PT), smart city utility (SCU), and technology proficiency (TP), have a very close functional alignment. We also verified the correlations between these items and found strong correlations. The factor analysis shows that 3 items are components of the variable smart city utility, 6 items are components of the variable perceived trust, 4 items are components of the variable technology proficiency, and 4 items are components of the variable technology proficiency.

\subsection{Structural equation modeling}

We test the hypothesized relationships using structural equation modeling (SEM) with R software. We used the SEM function in the lavaan package \citep{rosseel2012lavaan}. In our model, we use both categorical data and latent data, which restricts us to using composite measures for interactions. SEM was selected because of the ability to estimate direct and indirect effects  \citep{asparouhov2005sampling}. Since our data in the survey is Likert-scale, which is considered ordinal \citep{allen2007likert}, we adjusted the SEM method to ordinal items \citep{awang2016likert}. We used the Weighted Least Squares method (WLS) for estimating structural equation models that can handle categorical, ordinal, and continuous outcomes \cite{asparouhov2005sampling}.

Our model can be described by 4 equations in the model, whereby 4 exogenous variables have direct relations to the adoption of smart cities in the first model, and 3 exogenous variables have direct relations to the adoption of smart cities in the second model. It is important to note that the neighborhood variable is a categorical variable tested in the model as 4 dummy binary variables. To maintain one degree of freedom in the model, we examined 3 neighborhoods in the equations. Based on these assumptions, the actual equations of the model are:

\begin{flalign*}
Y_{DSU} =  \sum_{}^{}\alpha_{i}X_{n}+\beta X_{TP}+\varepsilon_{error} 
\\
Y_{PC} =  \sum_{}^{}\alpha_{i}X_{n}+\beta X_{TP}+\varepsilon_{error} 
\\
Y_{TP} =  \sum_{}^{}\alpha_{i}X_{n}+\varepsilon_{error}
\\
Y_{AoMDS} =  \sigma_{1}Y_{DSU}+\sigma_{2}Y_{TP}+\sigma_{3}Y_{PC}+\varepsilon_{error} 
\end{flalign*}
The structural equation model examines the interrelationships between the constructs:
Technology Proficiency (TP), Digital Service Utility (DSU), Perceived Privacy (PP), and Adoption of Municipal Digital Services (AoMDS), where $n$ represents the neighborhood.

\section{Results}

\subsection{Model Validation}

\begin{figure}[!h]
\includegraphics[width=\linewidth]{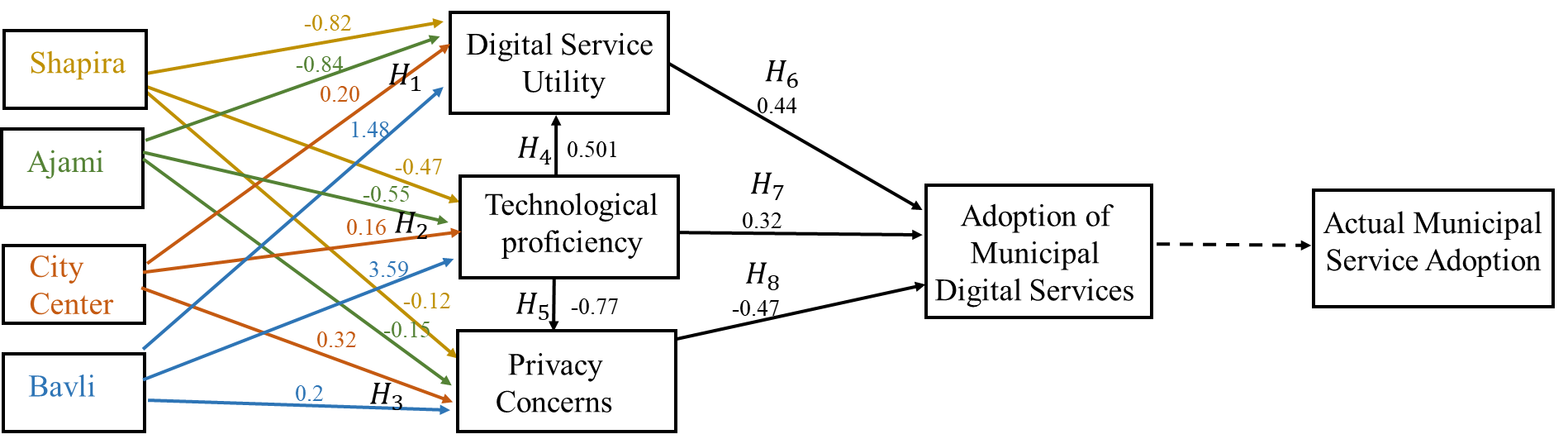}
\caption{Model results with SEM implementation}
\label{fig:validated_model}
\end{figure}

To evaluate whether neighborhoods explain the data better than demographics, we tested two structural equation models, as presented in Table \ref{tab:fit}. The first model includes neighborhoods as predictors, while the second model incorporates demographic variables. Fit indices such as Chi-Square, RMSEA, CFI, and TLI were compared to assess the relative explanatory power of each model. The neighborhood model demonstrated superior fit, with all indices meeting or exceeding established thresholds. Specifically, the Chi-Square test for the neighborhood model (p = 0.294) indicates an excellent fit under the null hypothesis, unlike the demographic model, which fails this test (p = 0.009). Moreover, the Tucker-Lewis Index (TLI) for the neighborhood model is 0.982, far surpassing the demographic model’s TLI of 0.789, suggesting that neighborhoods explain the data more effectively. Therefore, our findings are based on the neighborhood, which is described in Figure \ref{fig:validated_model}. 

\begin{table}[!h]
\caption{Fit indices for the two SEM models}
\begin{tabular}{|p{7cm}|p{2cm}|p{2cm}|p{2cm}|}
\hline
Fit measures                                                                            & Thresholds & Model witg neighborhoods & Model with demographics \\ \hline
\begin{tabular}[c]{@{}l@{}}Chi-square ($\chi^2$)\\ $H_0$ The model fits perfectly\end{tabular} & $p > 0.05$   & 0.294                       & 0.009                      \\ \hline
Degree of Freedom             &                & 5      & 118    \\ \hline
Chi-Square per Degrees of Freedom $\chi^2/DF$ & $\leq 2.0$         & 0.0541 & 0.0001 \\ \hline
Comparative fir index (CFI) & $\geq 0.90$         & 0.984  & 0.948  \\ \hline
RMSEA                         & $ < 0.06$ & 0.027  & 0.044  \\ \hline
Tucker-Lewis Index (TLI)      & $\geq 0.90$       & 0.982  & 0.789  \\ \hline
Standardized Root Mean Square Residual (SRMR)                                           & $< 0.08$   & 0.019                       & 0.015                      \\ \hline
\end{tabular}
\label{tab:fit}
\end{table}
In examining hypothesis 1, residency in the northern neighborhoods is positively associated with the digital service utility (DSU), and the southern neighborhoods are negatively associated with the smart city utility. The path coefficients of the northern neighborhoods are positive ($\beta$\textsubscript{CityCenter }$=$0.201, p<.05, $\alpha$\textsubscript{Bavli}$=$1.472, p<.05) and those for the southern neighborhood are negative ($\beta$\textsubscript{Ajami }$=$ -0.841, p<.05, $\beta$\textsubscript{Shapira}$=$ -0.836, p<.05). The relation of the City Center neighborhood is positive but not significant; we assume that this happens as a result of the residents from Bavli, who yield higher utility uses and who have a higher socioeconomic status than residents of other neighborhoods. The assumption for the southern neighborhoods is significantly negatively related to the digital service utility. 

The model supports hypothesis 2, showing that neighborhoods relate to technological proficiency (TP). Our results support this assertion for the southern neighborhoods ($\beta$\textsubscript{Ajami }$=$ -0.812, p<.10, $\beta$\textsubscript{Shapira}$=$ -0.466, p<.05) and for the northern neighborhoods, we obtain positive coefficients ($\beta$\textsubscript{CityCenter }$=$ 0.16, p<.05, $\alpha$\textsubscript{Bavli}$=$3.59, p<.05). Similarly, the model supports hypothesis 3 and neighborhoods are associated with privacy concerns. The results show that the variables are significant for our hypothesis ($\beta$\textsubscript{Ajami }$=$ -0.152, p<.05, $\beta$\textsubscript{Shapira}$=$ -0.116, p<.05, $\beta$\textsubscript{CityCenter }$=$ 0.324, p<.05 and $\alpha$\textsubscript{Bavli}$=$0.211, p<.05)

The results support hypothesis 4, which suggests a positive relationship between technological proficiency (TP) and digital service utility (DSU). TP is a significant and positive predictor of digital service utility ($\beta$\textsubscript{TP }$=$ 0.501, p<.05). This means that higher technological proficiency is associated with higher perceived utility of digital services. Similarly, we also hypothesized a positive relationship between technological proficiency (TP) and privacy concerns (hypothesis 4). The model confirmed a significant positive relation to privacy concerns (PC), which means that the higher the technological capabilities of a person, the more he or she will be concerned about his or her privacy and security preferences. 

For adoption of smart cities, we find in the model that TP, PT and DSU have significant relations to approach towards smart city adoption ($\beta$\textsubscript{TP }$=$ 0.328 p<.05, $\beta$\textsubscript{PC }$=$ -0.48, p<.05, $\beta$\textsubscript{DSU}$=$ 0.316, p<.05). As we assumed, privacy concerns (PC) has a negative relation to approach towards smart city because the more people are concerned about their information privacy and information security, the more difficult it is to adopt a smart city. Furthermore, TP and DSU have a significant positive relation to the approach towards smart cities, which means that technologically advanced people who use internet utilities more easily adopt the smart city. For the adoption of DigiTel, we find in the model that the approach towards smart city has a significant relation to the adoption of DigiTel ($\beta$\textsubscript{MDS}$=$ 0.651, p<.05)

Finally, we can show that 1) digital service utility (measured by DSU), 2) technological proficiency (measured by TP) and 3) privacy concerns (PC) as represented in our model are the critical factors for the adoption of approach towords smart city. Furthermore, the neighborhood as a latent variable influences those three variables, as we can say that the attitudes in the northern and southern neighborhoods differ.

The model shows us that residents from different neighborhoods perceived different factors at different levels, influencing their behavioral attitudes and intention to accept the smart city concept and use its services. As a result, adoption criteria for different residents with different levels of associations with technology can have significant implications for city planners and local authorities. Furthermore, in Appendix H, we can see the overall fit statistics for the model that, based on the recommended cutoffs, indicate a good fit.

\subsection{Adoption of DigiTel in Tel Aviv}
\paragraph{DigiTel Registration}
To provide external validity to our model results, we compared the self-reported adoption of DigiTel in our survey with two types of phenomenon: actual adoption of DigiTel from the Tel Aviv municipality reports and interaction with Tel Aviv's Facebook page. The DigiTel report from 2017 details the number of new residents registered for the service, divided according to city quarters. To test the homogeneity of proportions, we used the $\lambda$\textsuperscript{2}–test for equality of proportion. The result shows that there is sufficient evidence at the 0.05 (p-value$=$0.84) level to conclude that the two populations do not differ with respect to their DigiTel usability in each neighborhood.

\paragraph{Geo-located Facebook page interactions}
To obtain information idea of citizens’ interactions with the social networks of Tel Aviv, we examined the municipality's Facebook page and counted interactions that are geographically located. As part of the new communication services offered to the residents, they can contact the municipality through the city page by filling in a form, sending a private message, or posting a message in the visitor posts area. The visitor posts area can be used as an open-source resource and was used in our research. 

The information on Facebook posts was collected over six months (August 2016–January 2017). The information contains data on complaints by the residents on diverse issues and in different areas of the city. In this way, citizens can contact the municipality in a simple and accessible way and immediately refer to a problem that occurs in real time in a specific place in the city. Most of the posts contain information about the problem, the address, and pictures that illustrate what the problem is. The problems listed on the municipality page are diverse and include problems such as garbage removal, traffic hazards, and roadblocks. All the information on the complaints was collected and mapped onto (x,y) coordinates and are presented in Figure \ref{fig:tel_avivyafo_city_map_facebook}. 

To examine the relationship between our research results on the adoption of smart cities and the use of a Facebook page as one of the smart city technology platforms, we constructed a regression model. The regression model we offer predicts the number of complaints in our selected neighborhoods, depending on the smart city adoption variable and the socioeconomic status of the neighborhood. We used multiple regression analysis was used to test whether the adoption of a smart city and socioeconomic status significantly predicted number of complaints per resident in the neighborhoods. The regression model includes the following: 

\begin{equation}
C_n = \alpha \ast A_n+\beta \ast SE_n
\end{equation}
Where $N_r$ represents the number of complaints per resident in the neighborhood, $A_n$ represents the average adoption level per neighborhood according to our survey, and $SE_n$ is the socio-economic status of the neighborhood based on the statistical bourse data. We found that the variable of smart city adoption significantly predicted the number of Facebook complaints per resident in each neighborhood ($\alpha$ $=$ 6.9473, p<0.01), as did the socioeconomic status ($\beta$ $=$ -0.8632, p<.0.01). The result of the regression indicated that the two predictors explained 39.2$\%$ of the variance (R\textsuperscript{2 }$=$0.64, F(2,4) $=$ 6.944, p<0.01). The results further illustrate the connection between neighborhoods, citizens' acceptance of the smart city, and the usability of one of the most common platforms for contacting the municipality. This means that adopting the smart city by a citizen will make him or her use Facebook as a platform to connect with the municipality, but at the same time, if his or her socioeconomic status is lower, the use of the interface will decrease.

\section{Discussion}

\subsection{Neighborhood-Level Determinants of Smart City Adoption}

Our findings underscore the critical role of neighborhoods as determinants in adopting smart city technologies. Unlike traditional e-Government models that emphasize individual-level demographic factors such as age, income, and education \citep{shareef2011government, dirsehan2022smart}, this study highlights the spatial and social dimensions of urban living about the way residents interact with smart cities. Specifically, neighborhoods exhibit distinct patterns of technological proficiency, perceived utility, and privacy concerns, suggesting that the geographic and cultural context significantly shapes residents’ engagement with smart city initiatives. These results align with urban informatics research, which emphasizes the localized nature of digital experiences in urban environments \citep{hatuka2024conceptual, cranshaw2012livehoods}. 

Our results can quantify these differences in a much more nuanced way due to our representative sample that reflects populations often excluded from panel-based internet studies (such as the elderly population). Even when controlling for demographics and other variables, the differences are staggering: For example, technological proficiency in Bavli is more than 4 times higher than in Ajami. Similarly, Bavli’s perceived utility of smart city services is nearly 2.3 times higher than Ajami's. 

The study reveals a pronounced stratification in the adoption of smart city technologies linked to the socioeconomic profiles of neighborhoods. Northern neighborhoods in Tel Aviv, characterized by higher socioeconomic status, demonstrate higher technological proficiency and utility perceptions, whereas southern neighborhoods with lower socioeconomic status exhibit lower adoption levels. These findings expand upon the digital divide literature, which traditionally focuses on individual-level barriers, such as income and education, to include neighborhood-level constraints \citep{mossberger2007digital, hargittai2008digital}.

\subsection{Privacy Concerns Across Neighborhoods}

The findings of this study reveal significant differences in privacy perceptions across neighborhoods, with northern neighborhoods such as Bavli and City Center demonstrating higher privacy concerns compared to southern neighborhoods like Ajami and Shapira. Residents in Bavli, for example, exhibited a positive association with privacy concerns ($\alpha_{\text{Bavli}} = 0.211$, p < .05). In contrast, Ajami residents showed a negative association ($\beta_{\text{Ajami}} = -0.152$, p < .05). This disparity suggests that residents in affluent neighborhoods may have greater awareness of data privacy risks or higher expectations of data protection, aligning with prior research that links socioeconomic status to privacy awareness and agency \citep{kitchin2014real, gangadharan2017downside}. Conversely, lower privacy concerns in southern neighborhoods might stem from a lack of access to information, digital literacy, or trust in institutional safeguards, as highlighted by \citet{partridge2004developing}.

The study’s results expand the discourse on privacy in smart cities by demonstrating that privacy perceptions are not uniform across urban populations but are influenced by neighborhood-level factors. Previous studies, such as those by \citet{martinez2013pursuit} and \citet{elmaghraby2014cyber}, have predominantly focused on individual-level factors influencing privacy concerns, such as age, income, and education. By introducing the neighborhood as a significant factor, this research underscores the spatial dimension of privacy awareness. Residents in affluent neighborhoods like Bavli may exhibit heightened privacy concerns due to greater exposure to digital technologies and access to information about potential risks, as suggested by \citet{hargittai2008digital}. In contrast, southern neighborhoods like Shapira and Ajami may face structural barriers, including limited infrastructure and resources, which impede their ability to prioritize privacy concerns.

\subsection{Smart City Design and Policy Implications}
Our findings point to several implications for the design of smart city services. Municipal authorities and policymakers need to tailor smart city initiatives to the unique profiles of neighborhoods. For instance, neighborhoods with lower technological proficiency may benefit from targeted digital literacy programs. In contrast, neighborhoods with heightened privacy concerns may require enhanced trust-building mechanisms, such as transparency in data collection and robust privacy policies \citep{ragnedda2013digital, park2013digital}. By addressing these spatial disparities, cities can promote more equitable access to smart city services and mitigate the risk of digital exclusion.

This stratification has profound implications for urban governance and the design of smart city services. As \citet{sampson2008neighborhood} argues, neighborhoods are not merely geographic units but social constructs shaped by shared norms and collective resources. The study’s findings have implications beyond technology adoption, extending to broader urban management and governance practices. As \citet{sharkey2014and} highlight, the heterogeneity of urban experiences requires a nuanced approach to policymaking. Our study demonstrates that neighborhoods are not just passive recipients of smart city services but active agents shaping the adoption process through their unique social, cultural, and economic compositions.

Our findings have profound implications for policymakers aiming to address privacy concerns equitably. Smart city initiatives must prioritize privacy education campaigns tailored to the cultural and socioeconomic contexts of each neighborhood. For instance, workshops on data protection and digital literacy in neighborhoods like Ajami could help elevate awareness and build trust in municipal systems. Simultaneously, efforts to enhance transparency and accountability in data governance across the city can ensure that privacy concerns do not become a barrier to adoption in neighborhoods with heightened awareness.

This spatial disparity calls for a differentiated approach to privacy policy in smart cities. Municipalities must recognize that privacy concerns are not merely a function of individual demographics but are deeply intertwined with the socioeconomic and cultural fabric of neighborhoods. Strategies such as deploying multilingual, culturally sensitive communication and ensuring equitable access to secure technologies can help bridge the neighborhood gap. Furthermore, integrating participatory processes in privacy-related decision-making can empower residents from marginalized areas, aligning with the principles of inclusive urban governance.

This necessitates a shift from a one-size-fits-all approach to smart city implementation to a more granular, neighborhood-specific strategy. For instance, integrating participatory design processes that involve residents in decision-making can ensure that technologies are aligned with local needs and values \citep{caragliu2022smart}. Furthermore, the observed disparities in adoption suggest that smart city initiatives must be evaluated not only by their technological sophistication but also by their ability to promote social equity and inclusivity.

\section{Conclusions}

This study analyzes the critical role of neighborhood-level factors in shaping the adoption of smart city technologies, particularly in technological proficiency, perceived utility, and privacy concerns. By extending existing e-government adoption models to incorporate the spatial and social dimensions of urban environments, we demonstrate that the heterogeneity of neighborhoods significantly influences residents’ engagement with smart city initiatives. Northern neighborhoods, characterized by higher socioeconomic status and technological proficiency, exhibit greater adoption rates and heightened privacy concerns, while southern neighborhoods face barriers such as limited digital literacy and lower perceived utility of services. These findings emphasize the importance of tailoring smart city policies and interventions to the unique demographic, cultural, and socioeconomic profiles of neighborhoods. Addressing disparities through targeted digital literacy programs, privacy education, and equitable infrastructure investments is essential to fostering inclusive and effective smart city ecosystems. We hope that this research provides a foundation for urban planners and policymakers to design resident-centric smart city initiatives that bridge digital divides and enhance urban resilience.

\bibliographystyle{abbrvnat}

\bibliography{smart_bibliography}
\newpage
\appendix
\section{Appendix}

\renewcommand{\arraystretch}{1.5} % Adjust row spacing
\footnotesize
\begin{longtable}{|p{3cm}|p{12cm}|}
\caption{Survey items and constructs (translated from Hebrew)} \label{tab:items} 
\\ 
\hline
\textbf{Construct} & \textbf{Items} \\ \hline
\endfirsthead
\hline
\textbf{Construct} & \textbf{Items} \\ \hline
\endhead
\hline
\endfoot
Technology Proficiency (TP) & \begin{enumerate}[start=1]
    \item Do you use at least once a day your smartphone? (TP1)
    \item Do you use at least once a day your computer? (TP2)
    \item Have you ever downloaded apps to your smartphone? (TP3)
    \item Do you usually use the Internet from a computer in your home (not on the cellular phone)? (TP4)
\end{enumerate} \\ \hline

Digital Service Utility (DSU) & 
\begin{enumerate}[start=5]
    \item How often, if at all, do you shop online?
    \item When you use the Internet, how important is it for you to have the option to: 
    \begin{enumerate}
        \item Communications such as email or social network (DSU1)
        \item A civil organization such as signing petitions or organizing rallies (DSU2)
        \item Online interaction with public authorities, such as payments of bills (DSU3)
    \end{enumerate}
\end{enumerate} \\ \hline

Adoption of Municipal Digital Services (AoMDS) & 
\begin{enumerate}[start=7]
    \item How often, if at all, do you use digital municipal services such as Digi-Tel, payments, cultural events benefits, and school registration?
    \item Do you use digital municipal services for the following:
    \begin{enumerate}
        \item For payments
        \item For information such as blocking roads, city events (book week, shows)
        \item For benefits such as discounts for shows or parking lots, coupons, gifts
        \item For registration, for example, to the kindergartens, to sports races, to classes
        \item For using municipal WI-FI
    \end{enumerate}
    \item Are there any municipal digital services you avoid using?
    \item How much, in your opinion, do the digital municipal services improve the services offered to you as a citizen?

    \item Do you own a ‘Digi-Tel’ card? (DT)
    \item If you do, what are the reasons for it? 
    \begin{itemize}
        \item Benefits, comfort, don’t know, do not use
    \end{itemize}
    \item If you do not, what are the reasons for it? 
    \begin{itemize}
        \item Not aware, don’t need, not in your native language, privacy reasons
    \end{itemize}
\end{enumerate} \\ \hline

Smart City Attitude (SCA) & 
\begin{enumerate}[start=14]
    \item Have you heard of ‘The Smart City’?
    \item Is the idea of a smart city:
    \begin{enumerate}
        \item Making you hesitate? (SCA1)
        \item Causing you not to use the digital services? (SCA2)
        \item Making you enthusiastic about it? (SCA3)
    \end{enumerate}
\end{enumerate} \\ \hline

Privacy Concerns (PC) & 
\begin{enumerate}[start=16]
    \item Does it bother you that:
    \begin{enumerate}
        \item Your location info is being collected (PC1)
        \item The city takes pictures of public areas (PC2)
        \item Information about residents' consumption trends is collected (PC3)
        \item Information about kids' social activities is collected (PC4)
        \item Information is collected in general (PC5)
    \end{enumerate}
    \item Do you trust the municipality to protect its databases against hackers? (PC6)
\end{enumerate} \\ \hline

Demographics & 
\begin{enumerate}[start=18]
    \item How many years have you lived in the city?
    \item How many years have you lived in the neighborhood?
    \item How many people live in your home?
    \item How many rooms do you have at home?
    \item What is your occupation?
    \item How do you rate your family's financial situation?
    \item What is your marital status?
    \item What is your country of birth?
    \item What is your language?
    \item What is your total monthly income?
\end{enumerate} \\ \hline

\end{longtable}

\begin{longtable}{|l|l|l|l|l|}
\caption{Factor Loadings for Constructs} \label{tab:factor_loadings} \\ \hline
\textbf{Construct} & \textbf{Factor 1} & \textbf{Factor 2} & \textbf{Factor 3} & \textbf{Factor 4} \\ \hline
\endfirsthead
\hline
\textbf{Construct} & \textbf{Factor 1} & \textbf{Factor 2} & \textbf{Factor 3} & \textbf{Factor 4} \\ \hline
\endhead
\hline
\endfoot

DSU1  &            &            & 0.538      & 0.283      \\ \hline
DSU2  & 0.157      &            & 0.506      &            \\ \hline
DSU3  &            &            & 0.734      &            \\ \hline
PC1   & 0.538      & 0.102      & -0.212     & -0.120     \\ \hline
PC2   & 0.630      &            &            &            \\ \hline
PC3   & 0.688      &            &            &            \\ \hline
PC4   & 0.584      & 0.143      & -0.108     &            \\ \hline
PC5   & 0.771      &            & -0.123     &            \\ \hline
PC6   & 0.561      & -0.157     & -0.191     &            \\ \hline
TP1   &            & 0.570      &            &            \\ \hline
TP2   &            & 0.588      & -0.101     &            \\ \hline
TP3   &            & 0.522      &            & 0.866      \\ \hline
TP4   & 0.100      & 0.678      & -0.141     &            \\ \hline
SCA1  & 0.231      & -0.159     &            & -0.310     \\ \hline
SCA2  & 0.100      & -0.266     &            & -0.119     \\ \hline
SCA3  &            & -0.353     &            & -0.246     \\ \hline

\end{longtable}

\footnotesize
\begin{longtable}{|l|l|l|l|l|l|}
\caption{Demographics Validation Data by Neighborhood} \label{tab:demographics_validation} \\ \hline
\textbf{Neighborhood} & \textbf{Variable} & \textbf{Group} & \textbf{CBoC (\%)} & \textbf{Sample (\%)} & \textbf{p-value} \\ \hline
\endfirsthead
\hline
\textbf{Neighborhood} & \textbf{Variable} & \textbf{Group} & \textbf{CBoC (\%)} & \textbf{Sample (\%)} & \textbf{p-value} \\ \hline
\endhead
\hline
\endfoot

Ajami & Gender & Male   & 49\% (2850) & 41\% (51) & 0.065 \\ 
      &        & Female & 51\% (2930) & 59\% (73) &       \\ \cline{2-6}
      & Age    & 20-29  & 19\% (770)  & 19\% (24) & 0.983 \\ 
      &        & 30-64  & 62\% (2440) & 61\% (76) &       \\ 
      &        & 64+    & 19\% (740)  & 19\% (24) &       \\ \hline

Shapira & Gender & Male   & 52\% (3620) & 50\% (61) & 0.351 \\ 
        &        & Female & 48\% (3260) & 50\% (60) &       \\ \cline{2-6}
        & Age    & 20-29  & 18\% (920)  & 17\% (20) & 0.274 \\ 
        &        & 30-64  & 66\% (3480) & 65\% (79) &       \\ 
        &        & 64+    & 16\% (820)  & 18\% (22) &       \\ \hline

City Center & Gender & Male   & 50\% (7280) & 48\% (59) & 0.682 \\ 
            &        & Female & 50\% (7220) & 52\% (63) &       \\ \cline{2-6}
            & Age    & 20-29  & 21\% (2580) & 19\% (23) & 0.804 \\ 
            &        & 30-64  & 65\% (7860) & 66\% (81) &       \\ 
            &        & 64+    & 14\% (1700) & 15\% (18) &       \\ \hline

Bavli & Gender & Male   & 47\% (4400) & 41\% (51) & 0.207 \\ 
      &        & Female & 53\% (4940) & 59\% (72) &       \\ \cline{2-6}
      & Age    & 20-29  & 11\% (760)  & 11\% (14) & 0.976 \\ 
      &        & 30-64  & 62\% (4260) & 61\% (75) &       \\ 
      &        & 64+    & 27\% (1860) & 28\% (34) &       \\ \hline

\end{longtable}

\begin{table}[h]
\caption{Full SEM model}

\begin{tabular}{|p{2cm}p{2cm}|lll|}
\hline
\multicolumn{2}{|l|}{Variables}                                        & \multicolumn{3}{l|}{Adoption}                                                                   \\ \hline
\multicolumn{2}{|l|}{}                                                 & \multicolumn{1}{l|}{DSU}              & \multicolumn{1}{l|}{TP}               & PC              \\ \hline
\multicolumn{1}{|l|}{\multirow{4}{*}{Neighborhood}}   & Bavli          & \multicolumn{1}{l|}{1.472   (.0001)}  & \multicolumn{1}{l|}{3.596   (.0001)}  & 0.211   (.023)  \\ \cline{2-5} 
\multicolumn{1}{|l|}{}                                & City   Center  & \multicolumn{1}{l|}{0.201   (.308)}   & \multicolumn{1}{l|}{0.157   (.192)}   & 0.324   (.046)  \\ \cline{2-5} 
\multicolumn{1}{|l|}{}                                & Shapira        & \multicolumn{1}{l|}{-0.836   (.0001)} & \multicolumn{1}{l|}{-0.466   (.0001)} & -0.116   (.031) \\ \cline{2-5} 
\multicolumn{1}{|l|}{}                                & Ajami          & \multicolumn{1}{l|}{-0.841   (.0001)} & \multicolumn{1}{l|}{-0.812   (.0001)} & -0.152   (.044) \\ \hline
\multicolumn{2}{|l|}{Digital   Service Utility )DSU)}                  & \multicolumn{3}{l|}{\begin{tabular}[c]{@{}l@{}}AoMDS\\    \\  0.465  (.0001)\end{tabular}}      \\ \hline
\multicolumn{2}{|l|}{Technological   proficiency (TP)} &
  \multicolumn{1}{l|}{\begin{tabular}[c]{@{}l@{}}DSU\\    \\ 0.501   (.0001)\end{tabular}} &
  \multicolumn{1}{l|}{\begin{tabular}[c]{@{}l@{}}PT\\    \\ -0.770   (.0001)\end{tabular}} &
  \begin{tabular}[c]{@{}l@{}}MDS\\    \\ 0.328   (.0001)\end{tabular} \\ \hline
\multicolumn{2}{|l|}{Privacy   Concerns (PC)}                          & \multicolumn{3}{l|}{\begin{tabular}[c]{@{}l@{}}AoMDS\\    \\ -0.472    (.0001)\end{tabular}}    \\ \hline
\multicolumn{2}{|l|}{Adoption of Municipal Digital Services (AoMDS)} & \multicolumn{3}{l|}{0.65   (.0021)}                                                             \\ \hline
\end{tabular}
\end{table}

\begin{figure}[!ht]
\includegraphics[width=\linewidth]{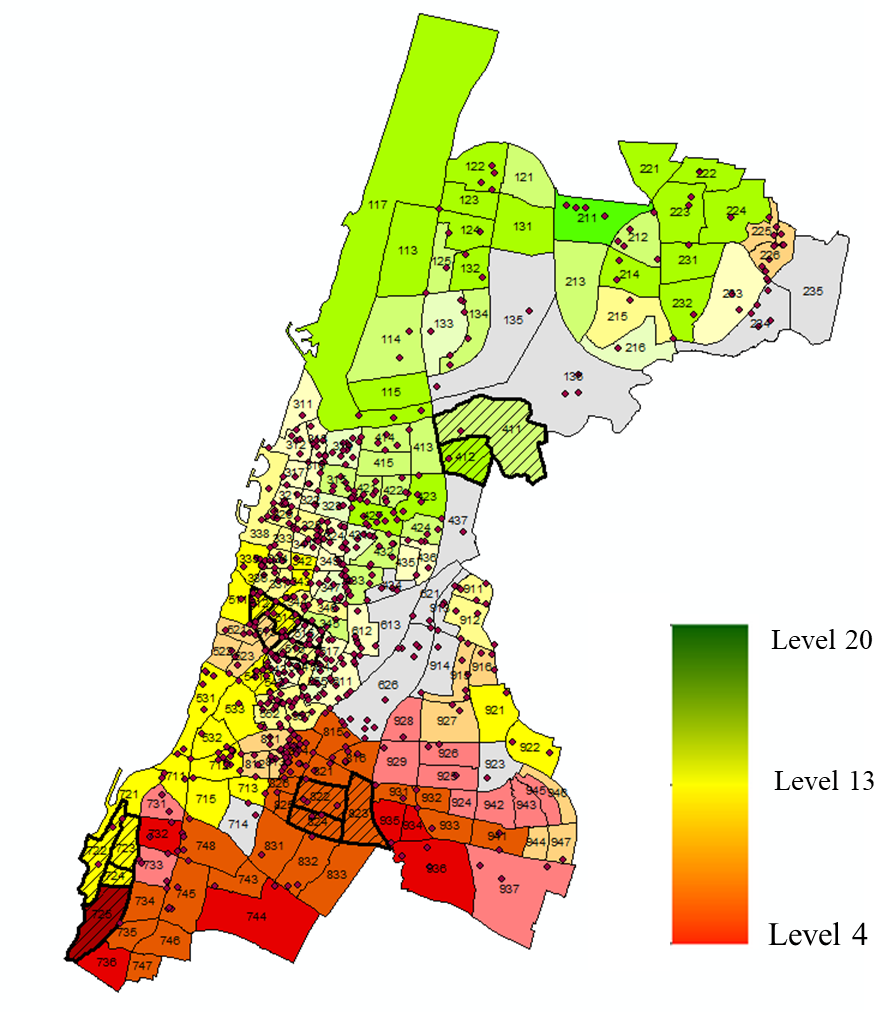}
\caption{Mapping the Facebook posts that include address data (red spots) in each statistical area of Tel Aviv.}
\label{fig:tel_avivyafo_city_map_facebook}
\end{figure}

\end{document}